\documentclass[prl,amsmath,amssymb,showpacs,showkeys,preprint,double-spaced]{revtex4}
\usepackage{graphicx}
\usepackage{dcolumn}
\usepackage{bm}
\usepackage{amssymb}
\usepackage{amsmath}
\begin{document}

\title{ Magnon BEC and Spin Superfluidity: a $^3$He primer}

\author{Yu.~M.~Bunkov\/\thanks{e-mail: yuriy.bunkov@grenoble.cnrs.fr}}

\address{ MCBT, Institut Neel, CNRS-UJF, 38042, Grenoble, France}

\author{G.~E.~Volovik}
\address{Low Temperature Laboratory, Helsinki University of
Technology, Finland\\
and L.D. Landau Institute for Theoretical Physics,
 Moscow, Russia}
\date{\today}

\begin{abstract}
Bose-Einstein condensation (BEC) is a quantum phenomenon of  formation of the
collective quantum state, in which the macroscopic number
of particles occupies the lowest energy state and thus is governed by a single wave function. Here we highlight the BEC in a magnetic subsystem --
the BEC of magnons,  elementary magnetic excitations. Magnon BEC is manifested
as the spontaneously emerging state of the precessing spins, in which all spins precess with the same frequency and phase even in the inhomogeneous magnetic field. We consider this phenomenon on example of spin precession in superfluid phases of $^3$He.
The magnon BEC in these phases has all the properties of spin superfluidity. The states of the phase-coherent precession belong to the class of the coherent quantum states, which manifest themselves by superfluidity, superconductivity, quantum Hall effect, Josephson effect and many other macroscopic quantum phenomena.
\bigskip

\end{abstract}


\maketitle

\section{Introduction}

Last decade was marked by the fundamental  studies  of mesoscopic
quantum states of dilute ultra cold atomic gases in the regime where the de Broglie
wavelength of the atoms is comparable with their spacing, giving
rise to the phenomenon of Bose-Einstein condensation (see, e.g.
Ref. \cite{revBEC}).
The formation of the Bose-Einstein condensate (BEC)
-- accumulation of the macroscopic number of particles in the lowest energy state --
was predicted by Einstein in 1925.
In ideal gas, all atoms are in the lowest energy state in the zero temperature limit. In dilute atomic gases, weak interactions between atoms produces a small fraction of the non-condensed atoms.
In the only known bosonic  liquid $^4$He which remains liquid at zero temperature, the BEC is strongly modified by interactions. The depletion of the condensate due to interactions is very strong: in the limit of zero temperature only about 10\%  of particles occupy the state with zero momentum.
Nevertheless, BEC still remains the key mechanism for the phenomenon of superfluidity in liquid
$^4$He: due to BEC  the whole liquid (100\% of $^4$He atoms) forms a coherent quantum state at $T=0$ and participates in the non-dissipative superfluid flow.

Superfluidity is a very general quantum property of  matter at low
temperatures, with variety of mechanisms and possible nondissipative superfluid
currents. These include supercurrent of electric charge in superconductors and
mass supercurrent in superfluid  $^3$He, where the mechanism
of superfluidity is the Cooper pairing;
hypercharge supercurrent in the vacuum of Standard Model of elementary
particle physics, which comes from the Higgs mechanism;
supercurrent of color charge in a dense quark matter in quantum chromo-dynamics; etc.
All these supercurrents have the same origin: the spontaneous breaking of the  $U(1)$
symmetry related to the conservation of the corresponding charge or particle
number, which leads to the so called off-diagonal long-range order.

Formally, the phenomenon of superfluidity requires the conservation of charge  or particle number.
However, the consideration can be extended to systems with a weakly violated conservation law, including a system
of sufficiently long-lived quasiparticles - discrete quanta of energy that can be treated as real particles in condensed matter. Here we shall consider the spin superfluidity -- superfluidity in the magnetic
subsystem of a condensed matter, which is represented by BEC of magnons - quanta of excitations of  the magnetic  subsystem, and is manifested as the spontaneous phase-coherent precession of spins first discovered in 1984 \cite{HPDexp,HPDtheory}.

 \subsection{BEC of quasiparticles}\label{BECQP}

At high temperatures, spins of atoms are in a disordered paramagnetic state,
which is similar to the high temperature phase of a weakly interacting gas.
With cooling the magnetic subsystem typically experiences a transition into an
ordered state, in which magnetic moments are correlated at long distances.
In cases when the magnetic $U(1)$ symmetry is spontaneously broken,
some people describe this phenomenon in terms of BEC of magnons \cite{mag1,mag2,mag3}.
Let us stress from the beginning that there is  the principal difference between the magnetic ordering and  the  BEC of quasiparticles which we are discussing in this review:

i. In some magnetic systems, the symmetry breaking phase transition starts when the system becomes softly unstable towards growth of one of the magnon modes. The condensation of this mode  leads finally to the formation of the true equilibrium ordered state.  In the same manner,  the Bose condensation of phonon modes may serve as a soft mechanism of formation of  the equilibrium solid crystals \cite{Kohn}. But this does not mean that the final crystal state is the  Bose condensate of phonons.
On the contrary,  BEC of quasiparticles is in principle a non-equilibrium phenomenon, since quasiparticles (magnons) have a finite life-time. In our case magnons live long enough to form a state very close to thermodynamic equilbrium BEC, but still it is not an equilibrium. In the final equilibrium state at $T=0$ all the magnons will die out.

ii. The ordered magnetic states are static equilibrium states which have diagonal long-range order. The magnon BEC is a dynamic state characterized by the off-diagonal long-range order, which is the main signature of spin superfluidity.

 \subsection{Non-conservation  vs coherence}\label{Nonconserv}

To prove that BEC of quasiparticles does really occur in a magnetic system, one should demonstrate the spontaneous emergence of coherence, and to show the consequences of the coherence: spin superfluidity, which in particular includes the observation of interference between two condensates.

We shall demonstrate that the finite life-time of magnons, and non-conservation of spin due to the spin-orbital coupling do not prevent the coherence of the magnon BEC. The gas of magnons can live a relatively long time, particularly at very low temperatures, sufficiently enough for formation of a coherent magnon condensate.  The non-conservation leads to a decrease of the number of magnons in the Bose gas until it disappears completely, but during this relaxation  the coherence  of BEC is preserved with all the signatures of  spin superfluidity: (i) spin supercurrent,  which transports the magnetization on a macroscopic distance more than 1 cm long; (ii) spin current Josephson effect which shows interference between two condensates; (iii) phase-slip processes at the critical current; (iv) spin current vortex  --  a topological defect which is an analog of a quantized vortex in superfluids, of an Abrikosov vortex in superconductors, and cosmic strings in relativistic theories; (v) Goldstone modes related to the broken $U(1)$ symmetry -- phonons in the spin-superfluid magnon gas; etc.

The losses of spin and energy in the magnon BEC can be compensated by pumping of additional magnons.  In this way one obtains a steady non-equilibrium state, which in our case is very close to the BEC in the thermodynamic equilibrium because the losses are relatively small.

\section{Coherent Larmor precession as magnon BEC}
\subsection{Disordered and coherent states of spin precession}

The magnetic subsystem which we discuss is the precessing magnetization. In a full correspondence with atomic systems, the precessing spins can be either in the normal state or in the ordered spin-superfluid state. In the normal state, spins of atoms are precessing with the local frequency determined by the local magnetic field and interactions. In the ordered state the precession of all spins is coherent: they spontaneously develop the common global frequency and the global phase of precession.

\begin{figure}[htt]
 \includegraphics[width=1.0\textwidth]{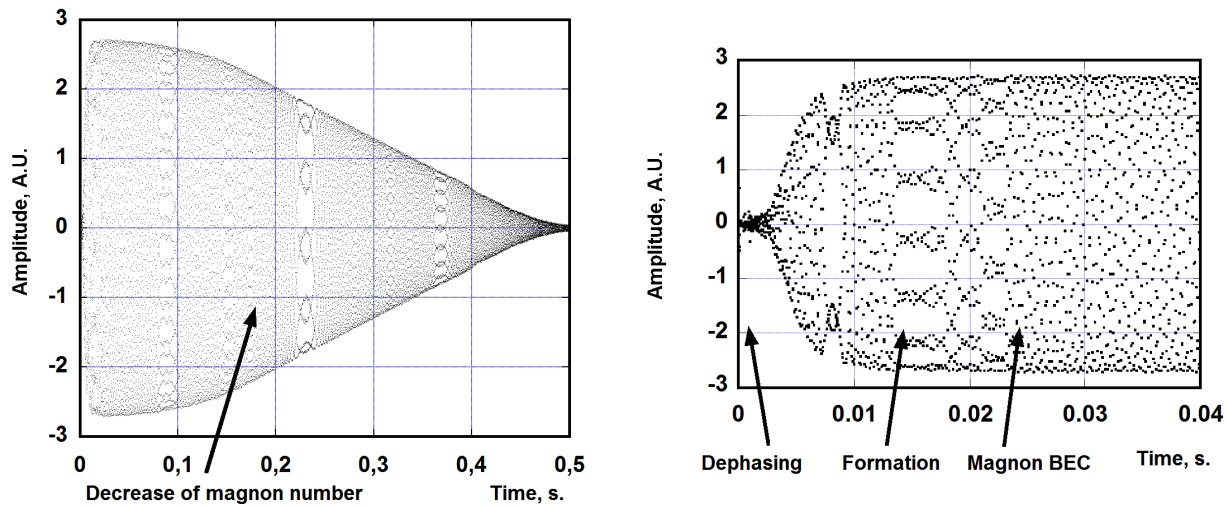}
 \caption{
 The stroboscopic record of  the induction decay signal on a frequency about 1 MHz.
{\it left}: During the first stage of about 0.002 s  the induction signal completely disappears due to dephasing. Then, during about 0.02 s, the spin supercurrent redistributes the magnetization and creates the phase coherent precession, which is equivalent to the magnon BEC state. Due to small magnetic relaxation, the number of magnons slowly decreases but the precession remains coherent.
  {\it right}: The initial part of the magnon BEC signal.
 }
 \label{amplitude}
\end{figure}

In NMR experiments the magnetization is created by an applied static magnetic field:
${\bf M}=\chi {\bf H}$, where $\chi$ is magnetic susceptibility. Then a pulse of the radio-frequency (RF) field ${\bf H}_{\rm RF}\perp  {\bf H}$ deflects the magnetization by an angle $\beta$, and after that the induction signal from the free precession is measured. In the state of the  disordered precession, spins almost immediately loose the information on the original common  phase and frequency induced by the RF field, and due to this decoherence the measured induction signal is very small.
In the ordered state, all spins precess coherently, which means that the whole macroscopic magnetization of the sample of volume $V$ is precessing
\begin{equation}
{\cal M}_x +i {\cal M}_y ={\cal M}_{\perp} e^{i\omega t+i\alpha} ~,~{\cal M}_{\perp}=\chi H V \sin\beta~.
\label{MagnetizationPrecession}
\end{equation}
This coherent precession is manifested as a huge and long-lived induction signal, Fig. \ref{amplitude}.

\subsection{Off-diagonal long-range order}

The superfluid atomic systems are characterized by the off-diagonal long-range order (ODLRO) \cite{Yang}.
In superfluid $^4$He and in the coherent atomic systems the operator
of annihilation of atoms with momentum ${\bf p}=0$ has a non-zero vacuum expectation value:
\begin{equation}
\left<\hat a_0 \right>={\cal N}_0^{1/2}  e^{i\mu
t+i\alpha}~,
\label{ODLRO}
\end{equation}
where ${\cal N}_0$ is the number of particles in the Bose condensate, which in the limit of weak interactions between the atoms  coincides at $T=0$ with the total number of atoms ${\cal N}$.

Eq. \eqref{MagnetizationPrecession} demonstrates that in the coherent precession the ODLRO is
manifested by a non-zero vacuum expectation value of the operator
of creation of spin:
\begin{equation}
\left<\hat S_+ \right>={\cal S}_x +i {\cal S}_y =\frac {{\cal M}_\perp}{\gamma} e^{i\omega
t+i\alpha}~,
\label{spinODLRO}
\end{equation}
where $\gamma$ is the gyromagnetic ratio, which relates magnetic moment and spin.
This analogy suggests that in the coherent spin precession the role of the particle number  ${\cal N}$ is played by the projection of the total spin on the direction of magnetic field ${\cal S}_z$. The corresponding symmetry group $U(1)$ in magnetic systems is the group of the $O(2)$ rotations about the direction of magnetic  field. This quantity ${\cal S}_z$ is conserved in the absence of the spin-orbit interactions.
The spin-orbit interactions transform the spin angular momentum of the magnetic subsystem to the orbital angular momentum, which causes the losses of spin ${\cal S}_z$ during the precession.  In our system of superfluid 3He, the spin-orbit coupling is relatively rather small,
and thus  ${\cal S}_z$ is quasi-conserved.  Because of the losses of spin the precession will finally decay, but during its long  life time the precession remains coherent, Fig. \ref{amplitude}.
This is similar to the non-conservation of the number of atoms in the laser traps, where the number of atoms decreases with time but this does not destroy the coherence of the atomic BEC.

The ODLRO in \eqref{spinODLRO} can be represented in terms of magnon condensation. To view that let us use the Holstein-Primakoff transformation, which relates the spin operators with the operators of creation and annihilation of magnons
\begin{eqnarray}
\hat a_0~\sqrt{1-\frac{\hbar a^\dagger_0 a_0}{2{\cal S}}}= \frac{\hat {\cal S}_+}{\sqrt{2{\cal S}\hbar}}~,
\label{MagnonAnnih}
\\
\hat a^\dagger_0~\sqrt{1-\frac{\hbar a^\dagger_0 a_0}{2{\cal S}}}=
\frac{\hat {\cal S}_-}{\sqrt{2{\cal S}\hbar}}~,
\label{MagnonCreation}
\\
\hat {\cal N}=\hat a^\dagger_0\hat a_0 =
\frac{{\cal S}-\hat{\cal S}_z}{\hbar}~.
\label{MagnonNumberOperator}
\end{eqnarray}
Eq. \eqref{MagnonNumberOperator} relates the number of magnons ${\cal N}$ to the deviation of spin ${\cal S}_z$ from its equilibrium value ${\cal S}_z^{({\rm equilibrium})}={\cal S}=\chi H V/\gamma$. In the full thermodynamic equilibrium,  magnons are absent. Each magnon has spin $-\hbar$,  and thus the total spin projection after pumping of ${\cal N}$ magnons into the system by the RF pulse is reduced by the number of magnons, ${\cal S}_z ={\cal S} - \hbar {\cal N}$. The ODLRO in magnon BEC is given by Eq. \eqref{ODLRO}, where ${\cal N}_0={\cal N}$ is the total number of magnons  \eqref{MagnonNumberOperator} in the BEC:
\begin{equation}
\left<\hat a_0 \right>={\cal N}^{1/2}  e^{i\omega
t+i\alpha}=\sqrt{\frac {2{\cal S}}{\hbar}} ~\sin\frac{\beta}{2} ~e^{i\omega
t+i\alpha}~,
\label{ODLROmagnon}
\end{equation}

Comparing \eqref{ODLROmagnon} and \eqref{ODLRO}, one can see that the role of the chemical potential in atomic systems $\mu$ is played by the global frequency of the coherent precession $\omega$, i.e. $\mu\equiv \omega$. This demonstrates that this analogy with the phenomenon of BEC in atomic gases takes place only for the dynamic states of a magnetic subsystem-- the states of precession. The ordered magnetic systems discussed in Refs. \cite{mag1,mag2,mag3} are static, and for them the chemical potential of magnons is always zero.

There are two approaches to study the thermodynamics of atomic systems:  one fixes either the particle number $N$ or the chemical potential $\mu$. For the magnon BEC, these two approaches correspond to two different experimental situations: the pulsed and continuous NMR, respectively. In the case of free precession after the pulse, the number of magnons pumped into the system is conserved (if one neglects  the losses of spin). This corresponds to the situation with the fixed ${\cal N}$, in which the system itself will choose the global frequency of the coherent precession (the magnon chemical potential). The opposite case is the continuous NMR, when a small RF field is continuously applied to compensate the losses. In this case the frequency of  precession is fixed by the frequency of the RF field, $\mu\equiv \omega = \omega_{\rm RF}$, and now the number of magnons will be adjusted to this frequency to match the resonance condition.

Finally let us mention that in the approach in which  ${\cal N}$ is strictly conserved and has quantized integer values,  the quantity $<\hat a_0> =0$ in Eg. (\ref{ODLRO}). In the same way the quantity $\left<\hat S_+ \right>=0$ in Eq. (\ref{spinODLRO}),
if spin ${\cal S}_z$ is strictly conserved and takes quantized values. This means that formally there is no precession if the system is in the quantum state with fixed spin quantum number ${\cal S}_z$. However, this does not lead to any paradox in the thermodynamic limit: in the limit of infinite ${\cal N}$ and ${\cal S}_z$,  the description in terms  of  the fixed ${\cal N}$ (or ${\cal S}_z$) is equivalent to the description in terms of with the fixed chemical potential $\mu$ (or frequency $\omega$).

\subsection{Order parameter and Gross-Pitaevskii equation}

As in the case of the atomic Bose condensates the main physics of the magnon BEC can be found from the consideration of the Gross-Pitaevskii equation for the complex order parameter.
The local order parameter is obtained by extension of Eq. \eqref{ODLROmagnon} to the inhomogeneous case and is determined as the vacuum expectation value of the magnon field operator:
\begin{equation}
\Psi({\bf r},t)=\left<\hat \Psi({\bf r},t)\right>~~,~~ n=\vert\Psi\vert^2~~,~~{\cal N}=\int d^3r ~\vert\Psi\vert^2~.\label{OrderParameter1}
\end{equation}
where $n$ is magnon density.
To avoid the confusion, let us mention that this order parameter \eqref{OrderParameter1} describes the coherent precession in any system, superfluid or non-superfluid. It has nothing to do with the multi-component order parameter which describes the superfluid phases of $^3$He \cite{VollhardtWoelfle}.

The corresponding Gross-Pitaevskii equation has the conventional form
$(\hbar=1)$:
  \begin{equation}
-i \frac{\partial \Psi}{\partial t}= \frac{\delta {\cal F}}{\delta \Psi^*}~,
\label{GP}
\end{equation}
where ${\cal F}\{ \Psi\}$ is the free energy functional.
In the coherent precession, the global frequency is constant in space and time (if dissipation is neglected)
\begin{equation}
\Psi({\bf r},t)=\Psi({\bf r})  e^{i\omega t}~,
\label{OrderParameter2}
\end{equation}
and the Gross-Pitaevskii equation transforms into the Ginzburg-Landau equation with $\omega=\mu$:
  \begin{equation}
 \frac{\delta {\cal F}}{\delta \Psi^*}- \mu\Psi=0~.
\label{GL}
\end{equation}
The Ginzburg-Landau  free energy functional has the following general form
  \begin{equation}
  {\cal F} -\mu {\cal N}=\int d^3r\left(\frac{\vert\nabla\Psi\vert^2}{2m_M} +(\omega_L({\bf r})-\omega)
  \vert\Psi\vert^2+F_{\rm so}(\vert\Psi\vert^2)\right),
  \label{GLfunctional}
\end{equation}
where $m_M$ is the magnon mass; and $\omega_L({\bf r})=\gamma H({\bf r})$  is the local Larmor frequency, which plays the role of external potential $U({\bf r})$ in atomic condensates.
Finally the last term $F_{\rm so}$ is analogous to the 4-th order term in the atomic BEC, which describes the interaction between the atoms of BEC.

\subsection{Spin-orbit  interaction as interaction between magnons}

In the magnetic subsystem of superfluid $^3$He, the interaction term in the Ginzburg-Landau  free energy is provided by the  spin-orbit interaction  -- interaction between the spin and orbital degrees of freedom.
Though the structure of  superfluid phases of $^3$He is rather complicated and is described by the multi-component superfluid order parameter  \cite{VollhardtWoelfle}, the only output needed for investigation of the coherent precession is the structure of the spin-orbit interaction, which is rather simple. The  spin-orbit interaction $F_{\rm so}$ depends on $\Psi$ and contains the second and the fourth order terms. Thus the spin-orbit interaction provides the effective interaction between magnons. It can be attractive and repulsive, depending on the orientation of spin and orbital orbital degrees with respect to each other and with respect to magnetic field. The orbital degrees of freedom  in  the superfluid phases of $^3$He are  characterized by the direction of the orbital momentum of the Cooper  pair  $\hat{\bf l}$, which also marks the axis of the spatial anisotropy of these superfluid liquids.
 By changing the orientation of $\hat{\bf l}$ with respect to magnetic field one is able to regulate the interaction term in experiments.

In superfluid $^3$He-B,  the spin-orbit interaction has a very peculiar properties.  After averaging over the fast precession of spins it acquires the following form \cite{BV}:
\begin{eqnarray}
  F_{\rm so}(s,l,\Phi)= \frac {2}{15} \frac {\chi}{\gamma^2} \Omega_L^2 [ (s l- \frac{1}{2} +
  \frac{1}{2}\cos{\Phi}(1+s)(1+l))^2+
 {\nonumber}
 \\
   \frac{1}{8}(1-s)^2(1-l)^2    + ( 1-s^2)(1-l^2)(1+\cos{\Phi})]
~.
  \label{FD}
  \end{eqnarray}
Here  $s=\cos\beta$, while  $l=\cos\beta_L=\hat{\bf l}\cdot \hat{\bf H}$ describes the orientation of the unit vector $\hat{\bf l}$ with respect to the direction $\hat{\bf H}$ of magnetic field.  The parameter $ \Omega_L$ is the so-called  Leggett frequency, which characterizes the magnitude of the spin-orbit interaction
and thus the shift of the resonance frequency  from the Larmor value caused by spin-orbit interaction. In typical experimental situations,  $\Omega_L^2\ll \omega^2$, which means that the frequency shift is relatively small. Finally $\Phi$ is another angle, which characterizes   the mutual orientation of spin and orbital degrees of freedom. It is a passive quantity: it takes the value corresponding to the minimum of $F_{\rm so}$ for given $s$ and $l$, i.e. $\Phi=\Phi(s,l)$.

To obtain $F_{\rm so}(\vert\Psi\vert^2)$ in \eqref{GLfunctional} at fixed $\hat{\bf l}$,  one must express $s$ via $|\Psi|$:
 \begin{equation}
1-s=1-\cos\beta = \frac{\hbar\vert\Psi\vert^2}{S}~,
\label{1-s}
\end{equation}
where $S=\chi H/\gamma$ is spin density. Since Eq. \eqref{FD} is quadratic in $s$, the spin-orbit interaction contains quadratic and quartic terms in $|\Psi|$. While the quadratic term modifies the potential $U$ in the Ginzburg-Landau free energy, the quartic terms describes the interaction between magnons.

\begin{figure}[htt]
 \includegraphics[width=0.6\textwidth]{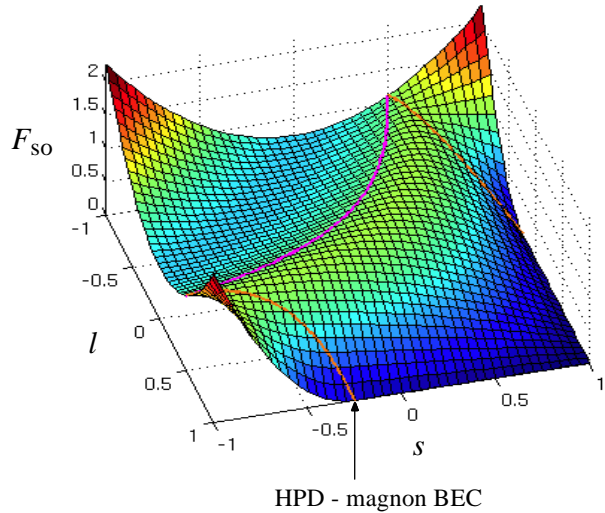}
 \caption{The profile of the spin-orbit energy as a function of $s=\cos\beta$ and orbital variable $l$, where
 $\beta$ is the tipping angle of precession and $l$ is the projection of the orbital angular momentum of a Cooper pair on the direction of magnetic field.
Spontaneous phase-coherent precession emerging at $l=1$ and $s\approx -1/4$ is called HPD
(Homogeneously Precessing Domain).}
 \label{profile}
\end{figure}

The profile of the spin-orbit interaction   $F_{\rm so}(s,l,\Phi(s,l))$ shown in Fig. \ref{profile} determines different states of coherent precession and thus different types  of magnon BEC in $^3$He-B, which depend on the orientation of the orbital vector $\hat{\bf l}$.
The most important of them, which has got the name Homogeneously Precessing Domain (HPD),
has been discovered 20 years ago \cite{HPDexp,HPDtheory}.

\section{Magnon BEC in $^3$He-B}

\subsection{HPD as unconventional magnon BEC}

In the right corner of Fig. \ref{profile}, the minimum of the free energy occurs for  $l=1$, i.e. for the orbital vector $\hat{\bf l}$ oriented along the magnetic field.   This means that if there is no other orientational effect on the orbital vector  $\hat{\bf l}$, the spin-orbit interaction orients it along the magnetic field, and one automatically obtains $l=\cos\beta_L=1$. The most surprising property emerging at such orientation is the existence of the completely flat region in Fig. \ref{profile}. The spin-orbit interaction is identically zero in the large range of the tipping angle $\beta$ of precession, for $1> s=\cos\beta >-\frac{1}{4}$:
   \begin{eqnarray}
F_{\rm so}(\beta, l=1)=0~~,~~{\rm when}~~\cos\beta> -\frac{1}{4}~,
   \label{SO1}
      \\
   F_{\rm so}(\beta,l=1)=\frac{8}{15}\frac {\chi}{\gamma^2}\Omega_L^2\left(\cos\beta
+\frac{1}{4}\right)^2 ~~,~~
{\rm when}~~\cos\beta< -\frac{1}{4}~.
 \label{SO2}
\end{eqnarray}
 Using Eq.~\eqref{1-s} one obtains the Ginzburg-Landau potential in \eqref{GLfunctional} with:
  \begin{eqnarray}
    F_{\rm so} \left(\vert\Psi\vert\right)=0~~,~~{\rm when}~~\vert\Psi\vert^2<n_c=\frac{5}{4}S~,
      \label{less104}
    \\
  F_{\rm so} \left(\vert\Psi\vert\right)=\frac{8}{15}\frac{\chi}{\gamma^2}\Omega_L^2 \left(
\frac{\vert\Psi\vert^2}{S}-\frac{5}{4}\right)^2 ~~,~~{\rm when}~~\vert\Psi\vert^2>n_c=\frac{5}{4}S~.
    \label{larger104}
\end{eqnarray}
 Eqs. \eqref{less104} and \eqref{larger104} demonstrate that when the orbital momentum is oriented along the magnetic field, magnons are non-interacting for all densities $n$ below the threshold value
  $n_c=(5/4)S/\hbar$. This is a really unconventional gas.

  \begin{figure}[htt]
 \includegraphics[width=0.6\textwidth]{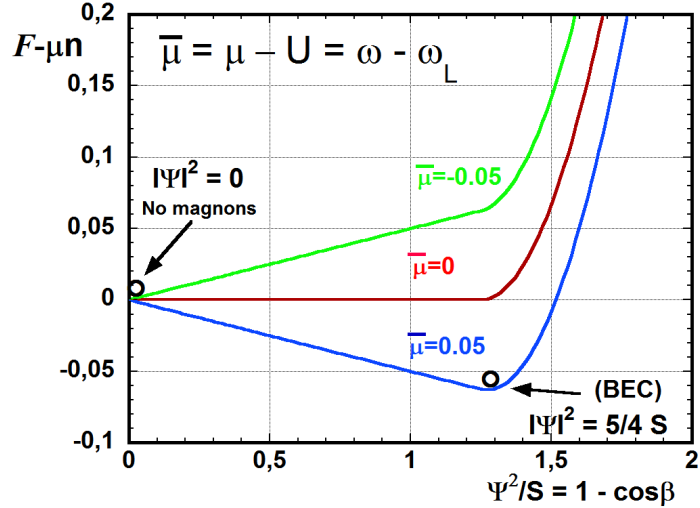}
 \caption{$F-\mu n$ for different values of the chemical potential
$\mu\equiv \omega$ in magnon BEC in $^3$He-B. For $\mu<U$, i.e. for $\omega<\omega_L$, the minimum of $F-\mu n$ corresponds to zero number of magnons, $n=0$. It is the static state  without precession. For $\mu=U$, i.e. for $\omega=\omega_L$, the energy is the same for all densities in the range $0\leq n \leq n_c$.
 For $\mu>U$, i.e. for $\omega>\omega_L$, the minimum of $F-\mu n$ corresponds
the magnon BEC with density $n\geq n_c$. This corresponds to the coherent precession of magnetization with tipping angle $\beta > 104^\circ$.}
 \label{F-muN}
\end{figure}

The energy profile of $F-\mu n$ is shown in Fig. \ref{F-muN} for different values of the chemical potential
$\mu\equiv \omega$. For $\mu$ below the external potential $U$, i.e. for $\omega<\omega_L$, the minimum of $F-\mu n$ corresponds to zero number of magnons, $n=0$. It is the static state of $^3$He-B without precession. For $\mu>U$, i.e. for $\omega>\omega_L$, the minimum of $F-\mu n$ corresponds
to the finite value of the magnon density:
 \begin{equation}
n=n_c \left(1 +\frac{3}{4} \frac{(\omega-\omega_L)\omega_L}{\Omega_L^2}\right)~.
\label{MagnonDensityBphase}
\end{equation}
This shows that the formation of HPD starts with the discontinuous jump to the finite density $n_c=5S/4\hbar$, which corresponds to coherent precession
with the large tipping angle  --  the so-called magic Leggett angle, $\beta_c\approx 104^\circ$ ($\cos\beta_c=-1/4$).
This is distinct from the standard Ginzburg-Landau energy functional (see Eq. \eqref{FDl=1} for magnon BEC in $^3$He-A-phase below), where the Bose condensate density smoothly starts growing from zero and is proportional to $\mu -U$  for $\mu>U$.

  \subsection{Two-domain precession}

    \begin{figure}[htt]
 \includegraphics[width=0.8\textwidth]{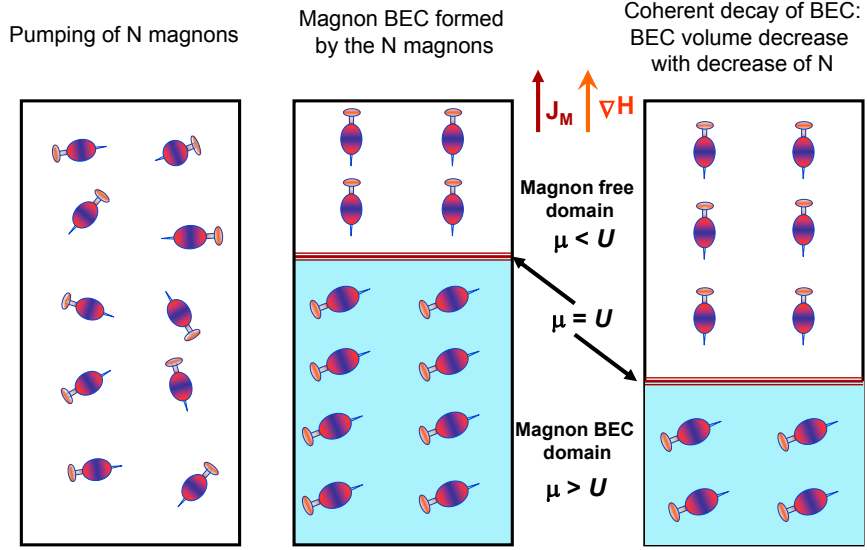}
 \caption{
 Two domains in the coherent precession in $^3$He-B. {\it left}: Incoherent spin precession after the pulse of the RF field deflects magnetization from its equilibrium value. The total number of magnons pumped into the system is ${\cal N}=({\cal S}-{\cal S}_z)/\hbar$ magnons. {\it middle}:
Formation of two domains. All ${\cal N}$ magnons  are concentrated in the lower part of the cell, forming the BEC state there. The volume of this domain is determined by the magnon density in BEC, $V={\cal N}/n$, where $n\approx n_c$. This volume determines the position $z_0$ of the domain boundary $z_0=V/A$, where $A$ is the area of the cross-section of the cylindrical cell. The position of the interface in turn determines the global frequency of presession, which is equal to the local Larmor frequency at the phase boundary, $\mu\equiv \omega=\omega_L(z_0)$.
{\it right} Decay of magnon BEC.  The number of magnons decreases due to spin and energy losses.
Since the magnon density in BEC is fixed (it is always close to $n_c$), the relaxation leads to the decrease of the volume of the BEC domain. However within this domain  the precession remains fully coherent. While the phase boundary slowly moves down the frequency of the global precession gradually decreases, Fig. \ref{frequency} {\it left}.
}
 \label{2Domain}
\end{figure}

The two states with zero and finite density of magnons, resemble the low-density gas state and the high density liquid state  of water, respectively. Gas and liquid can be separated in the gravitational field: the heavier  liquid state will be concentrated in the lower part of the vessel. For the magnon  BEC, the role of the gravitational field is played by the gradient of magnetic field:
\begin{equation}
\nabla U \equiv \nabla\omega_L= \gamma \nabla H~.
\label{FieldGradient}
\end{equation}
Thus applying the gradient of magnetic field along the axis $z$, one enforces phase separation, Fig. \ref{2Domain}.
The static thermodynamic equilibrium state is concentrated in the region of higher field, where
$\omega_L(z)>\omega$, i.e. $U(z)>\mu$. The magnon BEC -- the coherently  precessing state --  occupies the low-field region, where  $\omega_L(z)<\omega$, i.e. $U(z)<\mu$. This is the so-called Homogeneously Precessing Domain (HPD), in which all spins precess with the same frequency $\omega$ and the same phase $\alpha$.
In typical experiments the gradient is small,   and magnon density is close to the threshold value $n_c$.

The interface between the two domains is situated at the position $z_0$ where $\omega_L(z_0)=\omega$, i.e. $U(z_0)=\mu$. In the continuous NMR, the chemical potential is fixed by the frequency of the RF field:
$\mu=\omega_{\rm RF}$, this determines the  position of the interface in the experimental cel.

 In the pulsed NMR, the two-domain structure spontaneously  emerges after the magnetization is deflected by the RF pulse (Fig. \ref{2Domain}, {\it left} and {\it middle}). The position of the interface between the domains is determined by the number of magnons pumped into the system: ${\cal N}=({\cal S}-{\cal S}_z)/\hbar$. The number of magnons is quasi-conserved, i.e. it is well conserved during the time of the formation of the two-domain state of precession. That is why the volume of the domain occupied by the magnon BEC after its formation is $V={\cal N}/n_c$. This determined the position $z_0$ of the interface, and the chemical potential $\mu$ will be agjusted to this position: $\mu=\omega_L(z_0)$.

\begin{figure}[htt]
 \includegraphics[width=0.8\textwidth]{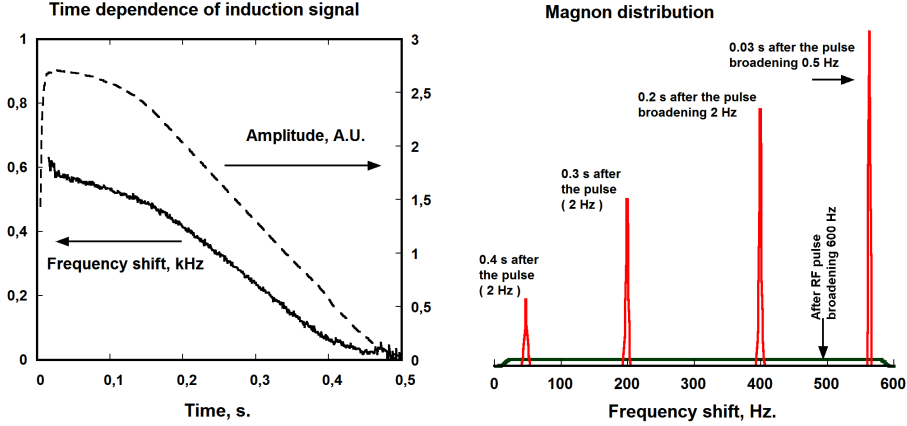}
 \caption{
The amplitude and frequency of the induction decay signal from magnon BEC.
{\it left}: The condensate occupies the domain where the chemical potential $\mu>U$ and radiates the signal
corresponding to the Larmor frequency at the domain
boundary of condensate. With relaxation the number of magnons
decreases, and the chemical potential moves to the region with a lower Larmor frequency.
{\it right}:  The spectroscopic distribution of magnons. Immediately after the RF
pulse each spin precesses with the local Larmore frequency. After  the BEC
formation, all the spins precess with the common frequency $\omega$ and
spontaneously emergent common phase $\alpha$. Due to relaxation the number of magnons decreases,
leading to  the continuously decreasing frequency. The small broadening of BEC state is
due to  relaxation. By comparing the initial broadening of the NMR line of about 600 Hz and
final broadening of about 0.5 Hz we can estimate that about 99.9 \% of the pumped magnons
are in the condensate.
}
 \label{frequency}
\end{figure}

In the absence of the RF field, i.e. without continuous pumping of magnons, the magnon BEC decays due to losses of spin. But the precessing domain (HPD) remains in the fully coherent Bose condensate state,  while the volume of the Bose condensate  gradually decreases due to losses and the domain boundary
slowly moves down  (Fig. \ref{2Domain}  {\it right}). The frequency $\omega$ of spontaneous coherence  as well as the phase of precession
remain   homogeneous across the whole Bose condensate domain,
but the magnitude of the frequency changes with time, since
it is determined by the Larmor frequency at the position of
the interface,
$\mu(t)\equiv \omega_L(z_0(t))$.  The change of
frequency during the decay is shown in Fig. \ref{frequency} {\it
left}.  This frequency change during the relaxation was the main
observational fact that led Fomin to construct the theory of the
two-domain precession \cite{HPDtheory}.

The details of formation of the magnon BEC are shown in Fig.
\ref{amplitude}, where the stroboscopic record of  induction decay
signal is shown. During the first stage of about 0.002 s  the
induction signal completely disappears due to dephasing. Then,
during about 0.02 s,  the phase coherent precession spontaneously
emerges, which is equivalent to the magnon BEC state. Due to a
weak magnetic relaxation, the number of magnons slowly decreases but
the precession remains coherent during the whole process of
relaxation. The time of formation of magnon BEC is essentially
shorter than the relaxation time, as clearly shown in  Fig.
\ref{amplitude} {\it right}.

\subsection{Other states of magnon BEC in  $^3$He-B}

Recent experiments allowed to probe the BEC states that emerge in the valley on the other side
of the energy barrier in Fig. \ref{profile}. This became possible  by  immersing the superfluid $^3$He in a very porous material called aerogel. By squeezing or stretching the  aerogel sample, one creates the global anisotropy which captures  the orbital vector $\hat{\bf l}$. This  allows to orient the orbital vector $\hat{\bf l}$ in the desirable direction with respect to magnetic field  \cite{JapGren,GrenB}.

For the transverse orientation of  $\hat{\bf l}$, i.e. for $l=0$, two new BEC states have been identified: one with attractive and the other with repulsive interaction between magnons. Omitting the constant term one obtains from Eq. \eqref{FD} the following spin-orbit interactions in these two states:
 \begin{eqnarray}
F_{\rm so} \left(\vert\Psi\vert, l=0\right)=-\frac{1}{4}\frac{\chi}{\gamma^2}\Omega_L^2
 \left(\frac{\vert\Psi\vert^2}{S}-\frac{4}{5}\right)^2
~~,~ ~ \vert\Psi\vert^2<\frac{S}{\hbar}~,
 \label{pot1}
   \\
F_{\rm so} \left(\vert\Psi\vert, l=0\right)=~\frac{1}{20}\frac{\chi}{\gamma^2}\Omega_L^2
\left(\frac{\vert\Psi\vert^2}{S}-2\right)^2
~~,~ ~ \vert\Psi\vert^2>\frac{S}{\hbar}~.
\label{pot2}
\end{eqnarray}

\subsection{Stability of coherent precession, spin supercurrent and phonons in magnon BEC}

There are several conditions required for the existence and stability of the magnon BEC. Some of them are the same as for the conventional atomic BEC, but there are also important differences.

i. The compressibility $\beta_M$ of the magnon gas must be positive:
\begin{equation}
\beta_M^{-1} = n\frac{dP}{dn}=  n^2\frac{d^2F}{dn^2} >0~.
\label{SpeedSound}
\end{equation}
This condition means that the fourth order term in the Ginzburg-Landau free energy should be positive, i.e.
the interaction between magnons should be repulsive.
The HPD in \eqref{larger104} at $n>n_c$, and the magnon BEC state in \eqref{pot2} satisfy this condition, while
the magnon BEC state in \eqref{pot1} is unstable.

ii. The compressibility of the magnon gas determines the speed of sound propagating in the magnon gas:
\begin{equation}
c_s^2= \frac{1}{m_M}\frac{dP}{dn}=  \frac{1}{nm_M\beta_M} ~.
\label{SpeedSound}
\end{equation}
In atomic superfluids, sound is the Goldstone mode of the spontaneously broken $U(1)$ symmetry. It is the consequence of a non-zero value of the  superfluid rigidity --  the superfluid density  $\rho_s$ which enters the non-dissipative supercurrent. The same takes place for magnon BEC, where the supercurrent carried by magnons  is  given by the traditional expression
\begin{equation}
{\bf J}= \rho_s {\bf v}_s~~,~~{\bf v}_s =\frac{\hbar}{m_M}\nabla\alpha~~,~~\rho_s(T=0)=nm_M~.
 \label{MassCurrent}
 \end{equation}
As in conventional superfluids, the superfluid density of the magnon liquid is determined by the magnon density $n$ and magnon mass $m_M$. To avoid the confusion let us mention that this superfluid density describes the coherent precession in magnetic subsystem and has nothing to do with the superfluid density of the underlying superfluid $^3$He.

Since each magnon carries spin $-\hbar$, the magnon mass supercurrent is accompanied by the magnetization supercurrent -- the supercurrent of the $z$-component of spin:
\begin{equation}
 {\bf J}_M=- \frac{\hbar}{m_M}~ {\bf J}=- n \frac{\hbar^2}{m_M}\nabla\alpha~.
  \label{SpinCurrent}
  \end{equation}
 The nonzero superfluid density $\rho_s>0$ is the main condition for superfluidity. For the HPD  in $^3$He-B this condition is fulfilled. The corresponding Goldstone phonon mode of the magnon BEC has been experimentally observed: it is manifested as twist oscillations of  the precessing domain in $^3$He-B   \cite{5}.

 As distinct from the conventional BEC,
 in magnon BEC one may introduce experimentally the symmetry breaking field which smoothly violates the $U(1)$ symmetry and induces a small gap (mass) in the phonon spectrum. This mass has been also measured, for details see Ref.  \cite{SBfield}.

iii. The next condition is applicable only to  the BEC of magnon quasiparticles, and is irrelevant for atomic condensates.  For quasiparticles the $U(1)$ symmetry is not strictly conserved. For magnetic subsystem, this is the $SO(2)$ symmetry with respect to spin rotations in the plane perpendicular to magnetic field, and it is violated by spin-orbit interactions.  The magnon BEC is a time dependent process, and it may experience instabilities which do not occur in equilibrium condensates of stable particles. In 1989 it was found that the original magnon condensate -- the HPD  state -- looses its stability below about 0.4 T$_c$ \cite{CatastrophExp} and experiences catastrophic relaxation. This phenomenon was left unexplained for a long time and only recently the reason was established: in the low-temperature regime, where dissipation becomes sufficiently small, the Suhl instability  destroys the homogeneous precession \cite{Catastroph}. This is the parametric instability, which leads to decay of HPD due to the parametric amplification of spin wave modes. It occurs because the spin-orbit interactions violate the
$U(1)$ symmetry.

\section{Magnon BEC in $^3$He-A}

As in the case of $^3$He-B, all the information on the $^3$He-A order parameter needed to study the coherent precession  is encoded in the spin-orbit interaction.
For $^3$He-A, the spin-orbit interaction has the form
\cite{BunkovVolovik1993}:
\begin{equation}
  F_{\rm so} \left(\vert\Psi\vert\right)= \frac {\chi\Omega_L^2}{4\gamma^2}
  \left[ -2\frac{\vert\Psi\vert^2}{S} +
  \frac{\vert\Psi\vert^4}{S^2}    +
    \left( -2+4 \frac{\vert\Psi\vert^2}{S}  -
  \frac {7}{4}\frac{\vert\Psi\vert^4}{S^2}\right)(1-l^2)\right]~.
  \label{FDA}
  \end{equation}
 In a static bulk $^3$He-A,
when $\Psi = 0$,  the spin-orbit energy $F_{\rm so}$ in  Eq.(\ref{FDA}) is minimized when the orbital vector $\hat{\bf l}$
is perpendicular to magnetic field, i.e. for $l= 0$. Then one has
\begin{equation}
  F_{\rm so} \left(\vert\Psi\vert, l=0\right)= \frac {\chi\Omega_L^2}{4\gamma^2}
  \left[-2+ 2\frac{\vert\Psi\vert^2}{S} -  \frac {3}{4}
  \frac{\vert\Psi\vert^4}{S^2}    \right],
  \label{FDl=0}
  \end{equation}
with a negative quartic term. The attractive interaction between
magnons destabilizes the BEC, which means that homogeneous
precession of magnetization in $^3$He-A becomes unstable, as was
predicted by Fomin \cite{Fomin1979} and observed experimentally in
Kapitza Institute \cite{InstabAB}.

\begin{figure}[htt]
 \includegraphics[width=0.6\textwidth]{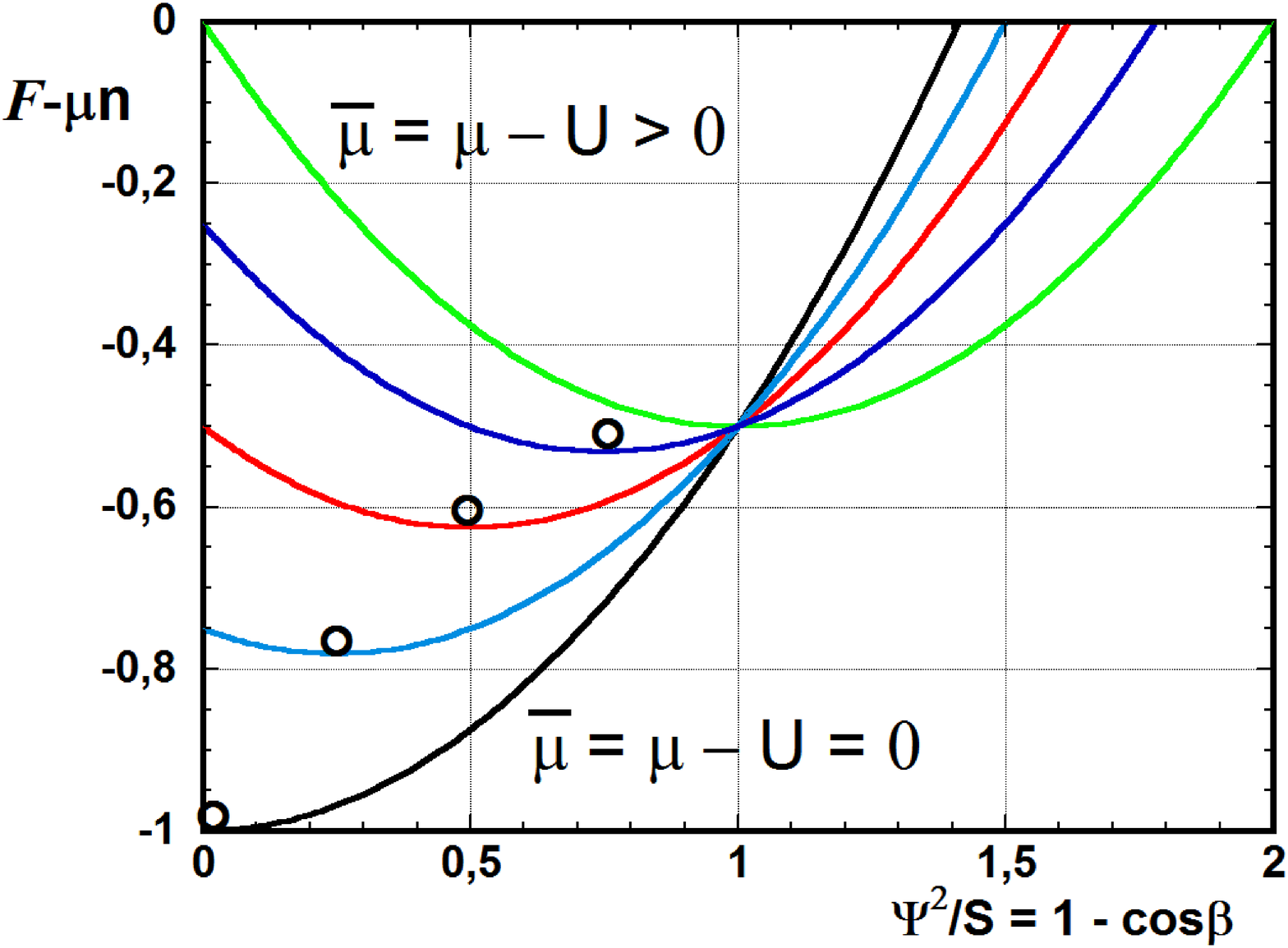}
 \caption{$F-\mu n$ for different values of the chemical potential
$\mu\geq U$ in magnon BEC in $^3$He-A. Magnon BEC in $^3$He-A is similar to BEC in atomic gases.}
 \label{3He-A}
\end{figure}

However, as follows from \eqref{FDA}, at sufficiently large magnon
density $n=\vert\Psi\vert^2$
 \begin{equation}
 \frac{8+\sqrt{8}}{7}~S> n> \frac{8-\sqrt{8}}{7}~S~,
\label{PositiveCondition}
\end{equation}
the factor in front of $l^2$ becomes negative. Therefore
it becomes energetically favorable to  orient the orbital momentum $\hat{\bf l}$
along the magnetic field, $l=1$.
For this  orientation one obtains the Ginzburg-Landau free energy with
\begin{equation}
  F_{\rm so} \left(\vert\Psi\vert, l=1\right)= \frac {\chi\Omega_L^2}{4\gamma^2}
  \left[ -2\frac{\vert\Psi\vert^2}{S} + \frac{\vert\Psi\vert^4}{S^2}  \right]~.
  \label{FDl=1}
  \end{equation}
It is similar to the  Ginzburg-Landau free energy for atomic BEC. The quadratic term modifies the potential $U$; the  quartic term is now positive.

In the language of BEC, this means that, with increasing the density
of Bose condensate, the originally attractive interaction between
magnons should spontaneously become repulsive when the critical
magnon density $n_c= S(8-\sqrt{8})/7$ is reached. If this happens, the magnon BEC becomes
stable and in this way the state with spontaneous coherent precession
could be formed \cite{BunkovVolovik1993}. However, such a self-sustaining BEC with
originally attractive magnon interaction has not yet been achieved
experimentally in bulk $^3$He-A, most probably because of large dissipation, due
to which the threshold value $n_c$ of the condensate density has not been
reached.

However, fixed orientation of $\hat{\bf l}$ with $l=1$ has been
prepared in $^3$He-A confined in a squeezed aerogel sample:
$\hat{\bf l}$ was oriented along the axis of uniaxially compressed
aerogel. For this geometry with $l=1$, the magnon BEC is stable, and
the first indication of coherent precession  in $^3$He-A has been
reported recently \cite{Sato2008,AphaseBEC}.  Contrary to the
unconventional magnon BEC in the form of HPD in $^3$He-B, the magnon
BEC emerging in the superfluid $^3$He-A is in one-to-one
correspondence with the atomic BEC, see  Fig. \ref{3He-A}.  For
$\mu>U$, the condensate density determined from equation $dF/dn=\mu$
continuously grows from zero  as $n \propto \mu -U$.

\section{Discussion}

\begin{figure}[htt]
 \includegraphics[width=0.8\textwidth]{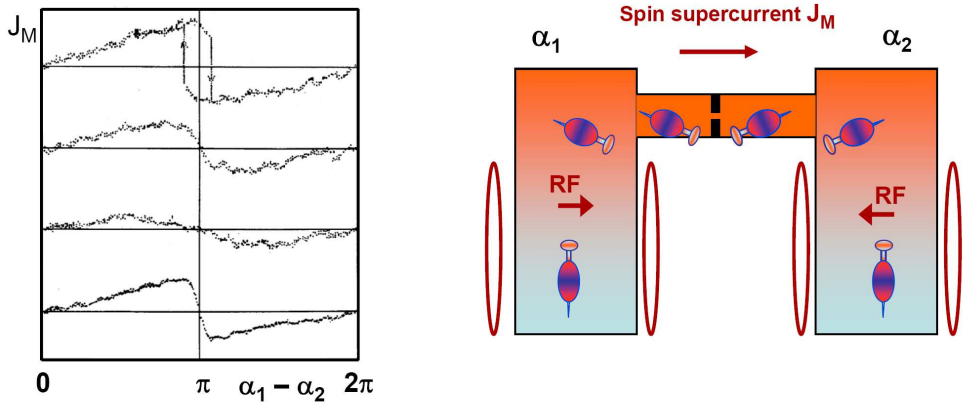}
 \caption{
 Josephson effect for magnon BEC in $^3$He-B which demonstrates the interference between two condensates. {\it left}: Spin current as a function of the phase difference across the junction,
 $\alpha_2-\alpha_1$, where $\alpha_1$ and  $\alpha_2$ are phases of precession in two coherently precessing domains connected by a small orifice ({\it right}).
Different experimental records correspond to a different ratio between the diameter of the orifice and the stiffness (magnetic coherence length) of the HPD state.
 }
 \label{Jos}
\end{figure}

The phase coherent precession of magnetizaton in superfluid $^3$He has all the
properties of the coherent Bose condensate of magnons. The main spin-superfluid properties of HPD have been verified already in the early experiments 20 years ago. These include spin supercurrent which transports the magnetization (analog of the mass current in
conventional superfluids); spin current Josephson effect and
phase-slip processes at the critical current \cite{6,7} (Fig. \ref{Jos}). Later on the spin
current vortex has been observed  \cite{Vortex} --  a topological
defect  which is an analog of a quantized vortex in superfluids
and of an Abrikosov vortex in superconductors.

The Bose condensation of magnons in superfluid $^3$He-B has many
practical applications. In Helsinki, owing to the extreme sensitivity  of the Bose
condensate  to textural inhomogeneity, the phenomenon of magnon BEC
has been applied to studies of
supercurrents and topological defects in $^3$He-B. The measurement
technique was called  HPD spectroscopy
\cite{HPDSpectroscopy,HPDSpectroscopy2}. In particular, HPD
spectroscopy provided a direct
experimental evidence for broken axial symmetry in the core of
a particular quantized vortex in $^3$He-B. Vortices with
broken symmetry in the core are condensed matter
analogs of the Witten  cosmic strings,  where the
addi\-tional $U(1)$ symmetry is broken inside the string core
(the so-called superconducting cosmic strings \cite{ewitten}).
The Goldstone mode of the vortex core resulting from the
spontaneous violation of rotational $U(1)$ symmetry in the core
has been observed
\cite{26}. The so-called spin-mass vortex,  which is a combined
defect serving as the termination line of the topological soliton
wall, has  also been observed and studied using  HPD spectroscopy
\cite{27}.

In Moscow \cite{Dmitriev}, Grenoble \cite{Grenoble1,Grenoble2}  and
Tokyo \cite{Tokyo1,Sato2008,Tokyo2}, HPD  spectroscopy proved to be extremely
useful for the investigation of the superfluid order parameter in
a novel system --  superfluid $^3$He  confined in aerogel.

There are many new physical phenomena related to the  Bose
condensation of magnons, which have been observed after the discovery of HPD  (see reviews \cite{BunkovHPDReview,V}). These include in particular compact objects with finite number of the Bose condensed magnons. At small number $N$ of the pumped magnons, the system is similar to the Bose condensate of the ultracold atoms in harmonic traps, while at larger $N$ the analog of the $Q$-ball in particle physics develops \cite{QBall}.

Magnon BEC is also possible in other magnetic systems, such as yttrium-iron garnet  films
\cite{Demokritov,Demidov}.
A very long lived induction signal  was observed in normal Fermi liquids: in spin-polarized $^3$He-$^4$He solutions
\cite{Nunes1992} and in normal liquid $^3$He \cite{Dmitriev1995}. It was explained as a coherently precessing  structure at the interface between the equilibrium domain and the domain with the reversed magnetization  \cite{Normal3He}. It would be interesting to treat this type of dynamic magnetic ordering as a new mode of magnon BEC.

 \section{Acknowledgement}

We are grateful to A. Golov, S. Demokritov,  M. Krusius  and T. Mizusaki  for illuminating discussions. This
work was done in the  framework of the project ``ULTIMA'' (NTO5-2-41909, Agence Nationale de
la Recherche), as the result of collaboration in the framework of the
Large Scale Installation Program ULTI of the European Union
(contract number: RITA-CT-2003-505313) and the collaboration between
CNRS and Russian Academy of Science, and was supported in part by the Russian Foundation
for Basic Research (grant 06--02--16002--a) and the
Khalatnikov--Starobinsky leading scientific school (grant 4899.2008.2).


\begin{thebibliography}{25}

\bibitem{revBEC} C.J. Pethick and H. Smith, "Bose-Einstein Condensation in Dilute Gases",
Ed. by Cambridge Univ. Press (2002); A.J. Leggett, Rev. Mod. Phys.
{\bf 73}, 307 (2001).

\bibitem{HPDexp} A.S. Borovik-Romanov, Yu.M. Bunkov, V.V. Dmitriev,  Yu.M. Mukharskiy, JETP Lett. {\bf 40},
1033, (1984);  JETP {\bf  61}, 1199 (1985).

\bibitem{HPDtheory}
I.A. Fomin, JETP Lett. {\bf 40}, 1036 (1984).

\bibitem{mag1} C. Ruegg,
Nature {\bf 423}, 63 (2003).

\bibitem{mag2} E. Della Torre,  L.H. Bennett and R.E. Watson,
{\it Phys.  Rev.  Lett.}, {\bf 94}, 147210 (2005).

\bibitem{mag3} T. Radu, H. Wilhelm, V. Yushankhai, D. Kovrizhin, R. Coldea, Z. Tylczynski, T. LŸhmann, and F. Steglich,
 {\it Phys.  Rev.  Lett.}, {\bf 95}, 127202 (2005).

\bibitem{Kohn} W. Kohn, D. Sherrington,
Rev. Mod. Phys. \textbf{42}, 1--11 (1970)

\bibitem{Yang}  C.N. Yang,
Rev. Mod. Phys. {\bf 34}, 694--704 (1962).


\bibitem{VollhardtWoelfle}  D. Vollhardt, P. W\"olfle, {\it The Superfluid Phases of Helium-3}, Taylor and Francis,. London 1990.

\bibitem{BV} Yu. M. Bunkov, G. E. Volovik,  {\it JETP}, {\bf 76}, 794 (1993).

\bibitem{JapGren} T. Kunimatsu, Kunimatsu T., Sato T., Izumina K., Matsubara A., Sasaki Y., Kubota M., Ishikawa O., Mizusaki T., Bunkov Yu.M.,
 JETP Lett. {\bf 86}, 216  (2007).

\bibitem{GrenB} J. Elbs, Yu. M. Bunkov, E. Collin, H. Godfrin, and G. E. Volovik
Phys. Rev. Lett. {\bf 100}, 215304 (2008)

 \bibitem{5} Yu.M. Bunkov,  V.V. Dmitriev, Yu.M. Mukharskiy,
JETP Lett.  {\bf 43}, 168 (1986)

 \bibitem{SBfield}  G.E. Volovik,
JETP Lett. {\bf 87}, 639--640 (2008).

\bibitem{CatastrophExp} Yu.M. Bunkov, V.V. Dmitriev, Yu.M. Mukharskiy, J. Nyeki, D.A. Sergatskov,
    Europhys. Lett. \textbf{8}, 645 (1989)

\bibitem{Catastroph} E.V. Sourovtsev, I.A. Fomin,
JETP Lett. \textbf{83}, 410 (2006);
Yu.M. Bunkov, V.S. Lvov, G.E. Volovik,
 JETP Lett. \textbf{83}, 530 (2006);
Yu.M. Bunkov, V.S. Lvov, G.E. Volovik,
JETP Lett. \textbf{84}, 289 (2006)


\bibitem{BunkovVolovik1993}  Yu.M. Bunkov and G.E. Volovik,
Europhys. Lett. {\bf 21}, 837--843 (1993).

\bibitem{Fomin1979} I.A. Fomin,
JETP Lett. {\bf 30}, 164--166 (1979).

 \bibitem{InstabAB} A.S.~Borovik-Romanov,
 Yu.M.~Bunkov, V.V.~Dmitriev, Yu.M.~Mukharskiy,
 JETP Lett.  {\bf 39}, 469--473 (1984).

\bibitem{Sato2008} T. Sato, T. Kunimatsu, K. Izumina,  A. Matsubara, M. Kubota, T. Mizusaki, Yu. M. Bunkov,
Phys. Rev. Lett. {\bf 101}, 055301 (2008).


\bibitem{AphaseBEC} Yu.M. Bunkov and G.E. Volovik,
Pis'ma ZhETF {\bf 89}, 356--361 (2009)

 \bibitem{6} A.S. Borovik-Romanov, Yu.M. Bunkov, V.V. Dmitriev,
Yu.M. Mukharskiy,
JETP Lett. \textbf{45}, 124 (1987)

 \bibitem{7} A.S. Borovik-Romanov, Yu.M. Bunkov, A. de Waard,
V.V. Dmit\-riev, V. Makrotsieva, Yu.M. Mukharskiy, D.A. Sergatskov,
JETP Lett. \textbf{47}, 478 (1988)


\bibitem{Vortex}
A.S. Borovik-Romanov, Yu.M. Bunkov, V.V. Dmitriev,  Yu.M. Mukharskiy,  D.A. Sergatskov, ÊÊ Ê Ê
Observation of Vortex-like Spin Supercurrent in $^3$He-B,
Physica  \textbf{B~165}, 649 (1990)

\bibitem{HPDSpectroscopy} Yu.M. Bunkov,
Physica \textbf{B~178}, 187 (1992)

 \bibitem{HPDSpectroscopy2} J.S. Korhonen,  Yu.M. Bunkov, V.V.
Dmitriev,  Y. Kondo, ÊM. Krusius, Yu.M. Mukharskiy, U. Parts, E.V.
Thuneberg,
Phys. Rev.  B \textbf{46},Ê13983--13990 (1992).

\bibitem{ewitten} E. Witten,
Nucl. Phys.  \textbf{ B249}, 557--592  (1985)

\bibitem{26} Y. Kondo, J.S. Korhonen, M. Krusius,  V.V.
Dmitriev, Y. M. Mukharsky, E.B. Sonin, G.E. Volovik,
Phys. Rev. Lett. \textbf{67}, 81--84 (1991)

\bibitem{27} Y. Kondo, J.S. Korhonen, M. Krusius,  V.V. Dmitriev, E.V. Thuneberg,  G.E. Volovik,
Phys. Rev. Lett. \textbf{68}, 3331--3334 (1992)

 \bibitem{Dmitriev} V.V. Dmitriev, V.V. Zavjalov,  D.E. Zmeev,  I.V. Kosarev, N.Mulders,
      JETP Lett. \textbf{76}, 321 (2002);
      V.V. Dmitriev, V.V. Zavjalov,  D.Ye. Zmeev,
    JETP Lett. \textbf{76}, 499 (2004)

 \bibitem{Grenoble1}   Yu.M. Bunkov, E. Collin, H. Godfrin,  R. Harakaly,
Physica \textbf{B~329}, 305 (2003)

 \bibitem{Grenoble2} J. Elbs, Yu. M. Bunkov, E. Collin, H. Godfrin,  G. E. Volovik,
Phys. Rev. Lett.  \textbf{100}, 215304 (2008)

\bibitem{Tokyo1}
 T. Kunimatsu, A. Matsubara,  K. Izumina, T. Sato,  M. Kubota, T. Takagi,
Yu.M. Bunkov, T. Mizusaki,
 J. Low Temp. Phys. \textbf{150}, 435--444 (2008)

 \bibitem{Tokyo2}
T. Kunimatsu, T. Sato, K. Izumina, A. Matsubara, Y. Sasaki, M. Kubota, O. Ishikawa, T. Mizusaki, Yu.M. Bunkov,
JETP Lett. \textbf{86}, 216--220 (2007)

\bibitem{BunkovHPDReview}  Yu.M.~Bunkov,
J. Low Temp. Phys. {\bf 135}, 337 (2004);
in: Progress in Low Temp.
Physics, ed. W. Halperin,  Elsevier (1995), vol. {\bf 14}, p.  69.

\bibitem{V} G.E. Volovik,
J. Low Temp. Phys. {\bf 153}, 266--284 (2008).


\bibitem{QBall} Yu.M. Bunkov, G.E. Volovik,
 Phys. Rev. Lett. \textbf{98}, 265302 (2007).

\bibitem{Demokritov}
S.O. Demokritov, V.E. Demidov, O. Dzyapko, G.A. Melkov, A.A. Serga, B. Hillebrands, A.N. Slavin,
Nature \textbf{443},  430--433 (2006).

 \bibitem{Demidov} V.E. Demidov,   O. Dzyapko,  S.O. Demokritov,  G.A. Melkov and A.N. Slavin,
 Phys. Rev. Lett. \textbf{100}, 047205 (2008).

 \bibitem{Nunes1992} G. Nunes, Jr., C. Jin, D.L. Hawthorne, A.M. Putnam, and D.M. Lee,
 Phys. Rev. B {\bf 46}, 9082--9103 (1992).

 \bibitem{Dmitriev1995} V.V. Dmitriev, S.R. Zakazov and V.V.  Moroz,
  JETP Lett. {\bf 61}, 309--315 (1995).

 \bibitem{Normal3He} Dmitriev V. V., Fomin I. A.,
 JETP Lett. {\bf 59}, 378--384 (1994).



\end{thebibliography}
\end{document}